\documentstyle[11pt,aas2pp4]{article}
\begin{document}
\tighten
\singlespace

\title{Observations of Small Scale ISM Structure in Dense Atomic Gas}
\author{J. T. Lauroesch\footnote{Visiting Astronomer,
	Kitt Peak National Observatory, National Optical
	Astronomy Observatories, which is operated by the Association of
	Universities for Research in Astronomy, Inc. (AURA) under cooperative
	agreement with the National Science Foundation.}}
\affil{Department of Physics and Astronomy}
\affil{Northwestern University}
\authoraddr{Evanston, Illinois 60208}
\authoremail{jtl@elvis.astro.nwu.edu}
\and
\setcounter{footnote}{0}
\author{David M. Meyer\footnotemark}
\affil{Department of Physics and Astronomy}
\affil{Northwestern University}
\authoraddr{Evanston, Illinois 60208}
\authoremail{meyer@elvis.astro.nwu.edu}

\begin{abstract}
We present high resolution (R$\sim$170,000) Kitt Peak National Observatory
C{\'{o}}ude Feed telescope observations of the interstellar
\ion{K}{1} 7698\AA\ line towards 5 multiple star systems with
saturated \ion{Na}{1} components.  We compare the \ion{K}{1} absorption
line profiles in each of the two (or three) lines of sight in these systems,
and find significant differences between the sight-lines in 3 out of the
5 cases.  We infer that the small scale structure traced by previous
\ion{Na}{1} observations is also present in at least some of the components
with saturated \ion{Na}{1} absorption lines, and thus the small scale
structures traced by the neutral species are occurring at some level in clouds
of all column densities.  We discuss the implications of that conclusion
and a potential explanation by density inhomogeneities.
\end{abstract}


\section{Introduction}

There is convincing evidence from recent radio and optical observations
that the diffuse interstellar medium (ISM) exhibits significant
subparsec-scale variations down to limits of a few AU (Frail {\it et al}.\ 
1994; Davis {\it et al}.\ 1996; Meyer \& Blades 1996; Watson \& Meyer 1996).
On somewhat larger scales, strong variations in interstellar \ion{Na}{1}
profiles are observed towards multiple late-type stars in globular clusters 
(Langer, Prosser, \& Sneden 1990; Bates et al.\ 1995).  In some cases,
the inferred densities of these structures far exceed the nominal diffuse
cloud values and approach those of molecular cloud cores.  For example,
Frail {\it et al}. (1994) found pervasive variations in the \ion{H}{1}
opacity on scales of 5-100 AU utilizing multiepoch observations of
21-cm absorption toward high velocity pulsars, which implied densities
of $n_H\sim10^4$--$10^5$$\rm cm^{-3}$.  It is not clear how such small,
dense structures can arise or be maintained in the low pressure
environments of diffuse clouds.

One difficulty with using the \ion{Na}{1} D lines as a tracer of
small scale structure is the increasing saturation of these lines
as one goes to larger \ion{H}{1} column densities.   Thus the bulk of
the gas in heavily reddened sightlines cannot be sampled using the D lines,
and one is generally limited to identifying variations in the weaker wings of
the line.  A potential probe of these sightlines is the
$\lambda$7698\AA\ interstellar \ion{K}{1} line. The lower cosmic abundance of K
coupled with differences in the photoionization cross section leads to
\ion{K}{1} columns that are typically a factor of $\approx$50 below those
of \ion{Na}{1} in diffuse interstellar clouds (Hobbs 1974).  Thus one can
use \ion{K}{1} to see if the small scale structure phenomenon extends from
the very diffuse components that have been well sampled using the \ion{Na}{1}
D lines (Watson \& Meyer 1996; Meyer \& Blades 1996) to highly reddened
sightlines such as HD 206267 (this paper, $E(B-V)$ $=$ 0.50).

In this {\it Letter}, we present Kitt Peak National Observatory (KPNO)
C\'{o}ude Feed echelle observations of the interstellar \ion{K}{1} absorption
towards members of 5 multiple star systems which have saturated
interstellar \ion{Na}{1} profiles.  These observations allow us to trace
the small scale variations previously observed using \ion{Na}{1} to much
more heavily reddened lines of sight. In addition, we present a simple
method for estimating the possible differences in density in the clouds
using the observed \ion{K}{1} column variations.

\section{Observations}

The observations were obtained in June 1998 with the
0.9m C\'{o}ude Feed telescope and spectrograph at Kitt Peak National
Observatory using camera 6 in echelle mode and a Ford 3000 x 1000
pixel CCD chip.  The resolution of the data was measured using the
ThAr lamp emission lines and is $\approx 1.75$~$\rm km s^{-1}$ at
the location of the \ion{K}{1}
$\lambda 7698.974$\AA\ line.  A total of 5 systems were observed: 4 binaries
or common proper motion doubles and one triple system (see Table~1).
Observations of the stars $\alpha$ Aql and $\alpha$ Leo were obtained
as a template for dividing out telluric absorption in the vicinity of
the \ion{K}{1} 7698\AA\ line.  For the majority of the stars, exposures
were taken on different nights at different grating tilts to reduce the
effect of any flaws in the CCD chip.  However, due to poor weather
conditions observations of $\rho$ Oph were limited to a single night
and only a single exposure was obtained for HD 206267 D.

Reduction of the data was done with the NOAO IRAF
\footnote{IRAF is distributed by the National Optical Astronomy
Observatories, which are operated by the Association of Universities for
Research in Astronomy, Inc., under cooperative agreement with the National
Science Foundation.} {\tt echelle} data reduction package.
The individual frames were first bias-subtracted and flat-fielded,
then the scattered light was removed.  For all stars except $\rho$ Oph
the individual orders were then extracted
and cleaned of cosmic ray hits.  Since both members of the binary $\rho$ Oph
were in the slit, only the portions of the combined stellar profile that were
dominated by either one of the two stars were extracted.  All of the
resulting one-dimensional spectra were then wavelength calibrated,
shifted to heliocentric coordinates, summed, and finally continuum fitted
using low order polynomials.  Figures~1 and 2 show the final \ion{K}{1}
line profiles.  The resulting S/N-ratios for these observations varies
widely, from as low as $\sim$20 for the faint star HD 206267~D to as
high as $\sim 130$ for $\beta^2$ Sco.  Column densities, b-values, and
relative velocities of the various components were derived by profile
fitting with the programs {\tt xvoigt} (Mar \& Bailey 1995) and
{\tt fits6p} (Welty, Hobbs, \& York 1991).  The wavelength and oscillator
strength of the \ion{K}{1} 7698\AA\ line were taken from Morton (1991); the
inclusion of hyperfine splitting (Welty, Hobbs, \& Kulkarni 1994) in the
fits does not significantly alter the derived columns or b-values.
The column density, b-value, and velocity of the various components were
free parameters in the initial fits, while in the final fits the velocities of
the components were fixed since the velocity differences between members of
a system were smaller than the uncertainties.  The resulting column
densities are listed in Table~2; the listed uncertainties include contributions
due to signal fluctuations, continuum placement errors, and the errors induced
by uncertainty in the b-values.  It must be noted
that if there is unresolved component structure that the column density may
be higher than that listed.  In particular, higher resolution observations
of $\rho$ Oph A suggest that there are at least three distinct components
making up what is fit here as a single main component along this line of
sight (Welty 1998).

\section{Discussion}

As seen in Figures~1 and 2, significant differences in the absorption profile
were seen between the interstellar \ion{K}{1} lines towards the
stars in 3 of the 5 systems.  The detection of significant
variation in the profiles of the \ion{K}{1} within the saturated cores of
the (previously observed) \ion{Na}{1} profiles shows that the small
scale structures occur even in regions of high column density.
It should be noted that the detection of
widespread variation in \ion{Na}{1} and \ion{K}{1} in multiple components make
a circumstellar origin for the observed small scale structures significantly
less tenable.  For example, towards $\beta$ Sco variations in \ion{Na}{1}
are observed at $\rm v\sim -24$~$\rm km s^{-1}$ and in \ion{K}{1} in the main
component at $\rm v\sim -10$~$\rm km s^{-1}$, with the enhancements
in \ion{Na}{1}
and \ion{K}{1} columns occurring towards different members of the binary.
One also sees variations in multiple \ion{K}{1} components towards the HD 206267
system, but one must note that the HD 206267 system bears a striking resemblance
to the Trapezium (Abt 1986), and like the Trapezium is in a region of recent,
active star formation.  Thus much of the variation towards this system may
be in gas which is located relatively near the stars (although not
circumstellar), and may be the result of interactions between the surrounding
medium and stellar winds and/or associated with the \ion{H}{2} region
IC 1396 centered on HD 206267A (O'Dell {\it et al}. 1993). 
While large differences were not detected towards $\rho$ Oph (HD 147933/4) or
HD 161270/89 it is possible that significant variation could be hidden in
the strong (possibly saturated) cores of these lines.  In any case, the observed
\ion{K}{1} variations suggest that the small scale structures previously
identified using \ion{Na}{1} are truly ubiquitous, and occur even in highly
reddened sightlines.

We can use the observed variations in the \ion{K}{1} to infer the
corresponding \ion{Na}{1} variations for comparison with previous optical
studies of small scale structure.  Hobbs (1974) identified a relationship
between the total line of sight column densities of \ion{K}{1} and
\ion{Na}{1}:\hfil\break
\begin{center}
N(\ion{K}{1})$\sim 0.012~$N(\ion{Na}{1})$^{1.2}$,
\end{center}\hfil\break
\noindent
where N(\ion{K}{1}) and N(\ion{Na}{1}) are in units
of $\rm 10^{11}$$\rm  cm^{-2}$.
Assuming the above ``typical'' relationship between N(\ion{K}{1}) and
N(\ion{Na}{1}), we see that the observed \ion{K}{1} variations correspond
to \ion{Na}{1} variations as large as $\rm\sim 1.3\times 10^{13}$$\rm cm^{-2}$.
Thus small scale variations occur over 3 orders of magnitude in \ion{Na}{1}
column density (Meyer \& Blades 1996; Watson \& Meyer 1996; this paper),
and therefore represent more than a population of small, low column clouds.
Furthermore, if we assume the ``typical'' relationship between
N(\ion{K}{1}) and N(\ion{H}{1}) from Hobbs (1974), we can estimate the
density $\rm n_H (cm^{-3})$ in the structures responsible for the
observed profile variations using the binary separation by assuming that
variations in \ion{K}{1} trace variations in \ion{H}{1}.  Then the observed
N(\ion{K}{1}) differences imply $n_H>10^3$$\rm cm^{-3}$, similar to
the densities inferred for a number of components in previous \ion{Na}{1}
studies (Meyer \& Blades 1996; Watson \& Meyer 1996).  Such densities are
also only slightly less than the densities ($n_H\sim 10^4$--$10^5$$\rm cm^{-3}$)
inferred by Frail {\it et al}. (1994) in their study of 21-cm absorption
towards high velocity pulsars.

The lack of observable changes in the \ion{Zn}{2} column density in
components where the \ion{Na}{1} column varies towards the binary $\mu$ Cru
suggest that at least some fraction of the variation detected in the
\ion{Na}{1} studies is due to density, temperature, or ionization fraction
fluctuations, and thus not indicative of variations in the hydrogen column
density between the lines of sight (Lauroesch {\it et al}. 1998).  If we
assume that the observed \ion{K}{1} variations are due to similar
fluctuations, we can use the measured column density differences to
estimate (albeit somewhat crudely) the difference in density and temperature
in these structures.  The recombination rate for potassium is proportional to
$\rm\sim n_e\times T^{-0.7}$ (P\'{e}quignot \& Aldrovandi 1986); if we assume
a neutral ideal gas in pressure equilibrium then the recombination rate will go
roughly as $\rm\sim n_e\times n_H^{0.7}$.  Typically it is assumed that the
$\rm n_e/n_H$ ratio is roughly constant in neutral interstellar clouds, with
the dominant source of electrons being carbon atoms.  Under this assumption
we can then estimate required density contrast from the observed columns
since N(\ion{K}{1})$\rm\propto n_H^{1.7}$.  We must note that as
the density increases and/or the temperature decreases in these clouds there
will be a corresponding increase in the N(\ion{C}{1})/N(\ion{C}{2}) ratio,
and thus a corresponding decrease in the $\rm n_e/n_H$ ratio.  However, if
N(\ion{C}{1}) $<<$ N(\ion{C}{2}) in both clouds then the $\rm n_e/n_H$ ratio
will be roughly constant.  In any case, we will initially assume
$\rm n_e/n_H$ is constant, and then identify (if necessary) any cases
in which a correction is required.

Table~2 lists the column densities from our profile fitting analysis
of these lines of sight.  Based on the measured differences in the
column densities for the various components we have estimated the
density contrast $\rm\delta n_H$ assuming an ideal gas with a constant
$\rm n_e/n_H$, then:\hfil\break
\begin{center}
$\rm\delta n_H =[$N(\ion{K}{1})$_{strong}$/N(\ion{K}{1})$_{weak}$]$^{0.6}$
\end{center}\hfil\break
\noindent
where we have defined the strong and weak components such that
N(\ion{K}{1})$_{strong} >$N(\ion{K}{1})$_{weak}$.  Thus
all values of $\rm\delta n_H$ will be $\geq 1$.  Looking at Table~2,
we note that the estimated density differences are generally quite small,
generally less than a factor of 2.  This shows how relatively small density
and temperature fluctuations can give rise to large differences in the
column densities of neutral species.

There have been a number of simplifying assumptions made in the
preceding argument which may not be correct.  First, we have assumed
that the observed column density differences reflect density and
temperature fluctuations and not \ion{H}{1} column variations.  Despite
the lack of \ion{Zn}{2} column density variations along the lightly
reddened $\mu$ Crucis line of sight (Lauroesch {\it et al}. 1998), the radio
observations of pervasive variations in the 21 cm opacity towards pulsars and
extra--galactic radio sources imply that some (or even all) of these structures
may be associated with \ion{H}{1} column density variations.  In addition, these
structures may not be in pressure equilibrium with the surrounding medium --
since we have no information about the lifetimes of these structures we cannot
say whether they are stable structures or not.  We have also assumed that
the thermal gas pressure dominates in these structures, implicitly ignoring
any magnetic field effects.  Thus we may have underestimated the
magnitude of the density fluctuations in the sightlines.  On the other hand,
we have assumed that the $\rm n_e/n_H$ ratio is constant, another possibility
is that there are pervasive fluctuations in the electron density on
small scales or that the ratio decreases in the denser clouds due to the
recombination of carbon.  Only by directly measuring the densities
($\rm n_e$ and $\rm n_H$) in individual components towards multiple systems
through the use of diagnostics such as the \ion{C}{1} and \ion{C}{2} fine
structure lines can the nature of these structures be fully explored.  Such
ultraviolet observations will also enable the identification of any sightlines
with \ion{H}{1} column density enhancements similar to those inferred from radio
studies.  However, it must be noted that optical studies remain an efficient
probe of small scale structures, and can be used to trace such structures on
a variety of scales.

\acknowledgments
It is a pleasure to acknowledge the support of the staff of KPNO, especially
Daryl Willmarth.  We would also like to thank Ed Jenkins, Eric Sandquist,
Dan Welty, Don York, and the referee for their valuable comments and
suggestions.

\smallskip\smallskip
\begin{figure}[hb]
\plotfiddle{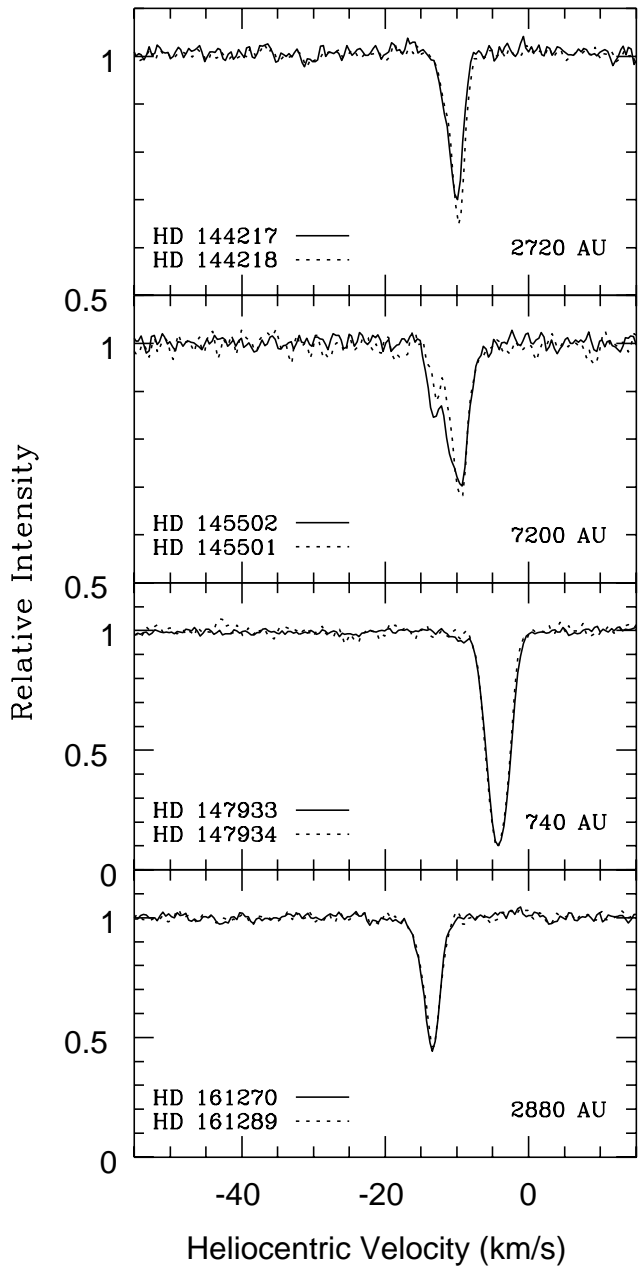}{6.2in}{0}{90}{90}{-155}{-155}
\caption{Observed interstellar \ion{K}{1} profiles for both members of the
binary star systems HD 144217/8 ($\beta$ Sco), HD 145502/1 ($\nu$ Sco),
HD 147933/4 ($\rho$ Oph), and HD 161270/89
taken using the KPNO C\'{o}ude Feed Telescope.  Note that the vertical
scale varies between plots.  There are significant differences in the
(relatively weak) profiles towards $\nu$ Sco and $\beta$ Sco, and a
lack of apparent variation in the strong (possibly saturated) lines
towards $\rho$ Oph and HD 161270/89.}
\end{figure}

\begin{figure}
\plotfiddle{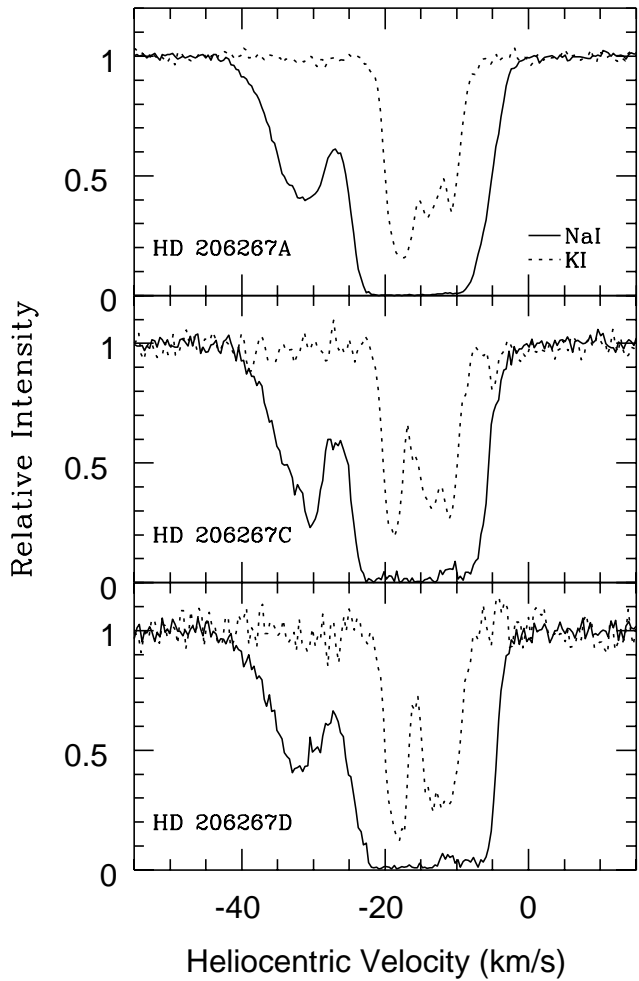}{4.5in}{0}{90}{90}{-155}{-155}
\caption{Comparison of the \ion{Na}{1} D$_2$ line profiles and
\ion{K}{1} profiles observed towards the HD 206267 multiple star system.
Note in particular the large number of variable components in the \ion{K}{1}
profile that are hidden within the broad saturated \ion{Na}{1} profile.
This shows the importance of \ion{K}{1} as a probe of small-scale structure
in regions of high column density.}
\end{figure}

\onecolumn

\begin{table}
\begin{center}
\caption{Stellar Data}
\smallskip
\tablewidth{0pt}
\begin{tabular}{lcccccc}
\tableline\tableline
 System & Alternate & V\tablenotemark{a} & E(B-V)\tablenotemark{b} &
	Separation\tablenotemark{a} & Separation\tablenotemark{c} \\
      & ID & (mag) & (mag)  & (\arcsec) & (AU)\\ \tableline
 HD 144217/8 & $\beta$ Sco & 2.6/4.9 & 0.21 & 13.6 & 2720 \\
 HD 145502/1 & $\nu$ Sco & 4.0/6.3 & 0.25 & 41.1 & 7200 \\
 HD 147933/4 & $\rho$ Oph & 5.0/5.9 & 0.48 & 3.2 & 740 \\
 HD 161270/89 &        & 6.2/6.6 & 0.07 & 20.6 & 2880 \\
 HD 206267~A/C/D &     & 5.6/8.4/8.0 & 0.52 & \tablenotemark{d} &
	\tablenotemark{d} \\ \tableline
\end{tabular}
\tablenotetext{a}{Yale Bright Star Catalog (Hoffleit \& Jaschek 1982).}
\tablenotetext{b}{Derived using the observed (B-V) and spectral
	type from the Yale Bright Star Catalog (Hoffleit \& Jaschek 1982),
	and the intrinsic colors from Mihalas \& Binney (1981)}
\tablenotetext{c}{Projected separation between primary and secondary(s)
	based on the spectroscopic parallax using the spectral types
	given in the Yale Bright Star Catalog (Hoffleit \& Jaschek 1982).}
\tablenotetext{d}{Triple system with separations of A--C 11.7\arcsec, 
	A--D 19.9\arcsec, and C--D 31.6\arcsec.  These angular
	separations correspond to projected separations of
	14,300 AU, 24,300 AU, and 38,550 AU respectively.}
\end{center}
\end{table}

\begin{table}
\begin{center}
\caption{Column Densities}
\smallskip
\tablewidth{0pt}
\begin{tabular}{lcccc}
\tableline\tableline
 System & Velocity & N(\ion{K}{1})\tablenotemark{a}
		& b-value\tablenotemark{b} & $\delta n_H$\tablenotemark{c} \\
        & ($\rm km s^{-1}$) & ($\rm\times 10^{10}\ cm^{-2}$) &
		($\rm km s^{-1}$) & ($\rm cm^{-3}$) \\ \tableline
 HD 144217/8 &  -9.9 & 12.3$\pm 1.0$/16.6$\pm 2.4$ & 0.4$\pm$0.1 & 1.2 \\
             & -11.7 & 3.3$\pm 0.6$/2.3$\pm 0.5$ & 0.4$\pm$0.1 & 1.2\\
 HD 145502/1 & -9.3  & 11.5$\pm$0.8/13.1$\pm 1.3$ & 0.7$\pm$0.1 & 1.1 \\
             & -11.1 & 7.1$\pm 1.0$/2.6$\pm 1.1$ & 0.3$\pm$0.1 & 1.8 \\
             & -13.3 & 6.1$\pm$1.0/4.4$\pm$2.3 & (0.2) & 1.2 \\
 HD 147933/4 & -7.8  & 117$\pm$18/118$\pm$22 & 1.3$\pm$0.1 & $<$1.3 \\
             & -12.5 & 1.7$\pm$0.5/$<$1.3 & 0.9$\pm$0.3 & $>$1.0 \\
 HD 161270/89 & -16.5 & 32.4$\pm$4.0/29.8$\pm$3.5 & 0.85$\pm$0.15 & $<$1.2 \\
 HD 206267~A/C/D & -18.9 & 53.5$\pm$4.7/86.8$\pm$10.6/55.3$\pm$12.6 &
			1.2$\pm$0.1 & 1.3 \\
                 & -17.2 & 45.5$\pm$6.5/$<$3.6/54.3$\pm$14.7 & 0.9$\pm$0.1 & $>$4.2 \\
                 & -16.0 & 20.5$\pm$5.5/20.5$\pm$8.2/$<$8.2 & (0.3) & $<$1.6 \\
                 & -14.5 & 64.8$\pm$10.7/47.1$\pm$12.6/64.8$\pm$36.4 & 0.3$\pm$0.1 &
			1.2 \\
                 & -13.0 & 28.2$\pm$2.7/36.0$\pm$7.6/48.5$\pm$12.5 &
			0.55$\pm$0.15 & 1.2 \\ 
                 & -10.8 & 45.2$\pm$3.7/66.7$\pm$2.4/83.1$\pm$18.2 &
			0.6$\pm$0.1 & 1.3 \\
                 & -8.9  & 6.9$\pm$1.7/4.2$\pm$2.6/6.9$\pm$3.5 &
			(0.2) & $<$2.7 \\
                 & -4.8  & $<$1.3/6.9$\pm$3.6/$<$2.0 &
			0.2$\pm$0.1 & $>$1.7 \\ \tableline
\end{tabular}
\tablenotetext{a}{Column densities with 1$\sigma$ errors (or 2$\sigma$ upper
	limits) which include the uncertainties due to signal fluctuations,
	continuum placement errors, and the errors induced by uncertainty in the
	b-values.  Values for the various members of the systems are separated by /'s.}
\tablenotetext{b}{Derived b-values, values in ()'s were fixed in
	the analysis and are less well determined.}
\tablenotetext{c}{Estimated density contrast (see text).  Note that
	for the HD 206267 system we give $\rm\delta n_H$ only for
	HD 206267~A and C due to the relatively large uncertainties
	in the fits to the lower S/N--ratio data for HD 206267~D.}
\end{center}
\end{table}

\end{document}